\documentclass[aps,prl,twocolumn,groupedaddress]{revtex4}

\usepackage{graphicx}
\usepackage{amssymb}

\newcommand{\rs}{\rm \scriptscriptstyle}

\begin{document}

\title{Atomic quantum simulator for lattice gauge theories and  
ring exchange models}

\author{H.P.\ B\"uchler$^{1,2}$}
\author{M.\ Hermele$^{3}$}
\author{S.D.\ Huber$^{4}$}
\author{Matthew P.A.\ Fisher$^{5}$}
\author{P. Zoller$^{1,2}$}

\affiliation{$^{1}$Institute for Quantum Optics and Quantum Information
of the Austrian Academy of Science, 6020 Innsbruck, Austria }
\affiliation{$^{2}$Institut f\"ur theoretische Physik, Universit\"at Innsbruck,
6020 Innsbruck, Austria}

\affiliation{$^{3}$Department of Physics, University of California, 
Santa Barbara, California 93106, USA}
\affiliation{$^{4}$Theoretische Physik, ETH-H\"onggerberg, CH-8093
Z\"urich, Switzerland}

 \affiliation{$^{5}$Kavli Institute for Theoretical Physics, University 
of California, Santa Barbara, California 93106, USA }

\date{\today}

\begin{abstract}

We present the design of a ring exchange interaction in cold atomic gases
subjected to an optical lattice using well understood tools for manipulating and
controlling such gases. 
The strength of this interaction can be tuned independently and describes the
correlated hopping of two bosons.
We discuss a setup
where this coupling term may allows for the realization and observation of
exotic quantum phases, including a deconfined insulator described by the Coulomb
phase  of a three-dimensional ${\rm U}(1)$ lattice gauge theory.

\end{abstract}


\maketitle

Loading cold atomic gases into optical lattices allows for the realization of
bosonic and fermionic Hubbard models, and offers the possibility for the
experimental study of strongly correlated systems within a highly tunable
environment.  Starting from the prediction of a superfluid to Mott-insulator
phase transition in bosonic atomic gases \cite{jaksch98} and the subsequent
observation of the Mott insulating phase \cite{greiner02,stoferle04,fertig04},
many new tools for manipulating and controlling quantum gases have been
developed \cite{jaksch04}.  In this Letter, we combine these tools in order to
drive the atomic gas with an additional ring exchange interaction. We identify a
promising system where this coupling may allow for the realization and
observation of an exotic quantum insulator \cite{hermele04,moessner03} described by the
Coulomb phase of a three-dimensional ${\rm U}(1)$ lattice gauge theory
\cite{kogut79};  in quantum magnetism this phase is known as a ${\rm U}(1)$ spin
liquid.

Recently, studies of boson models with large ring exchange have yielded
significant progress in the search for microscopic Hamiltonians exhibiting
exotic phases \cite{moessner01,balents02,motrunich02}.  This search has been the
focus of much effort in two-dimensional systems, due to potential relevance for
high-$T_c$ cuprates.  Some ring exchange models exhibit a local conservation
law, and can be mapped onto lattice gauge theories and often also quantum dimer
models (QDM) \cite{rokhsar88}.  A number of such models in two and three
dimensions have been shown to possess deconfined insulating ground states
\cite{senthil04-review}.  Many of the three-dimensional models, including those
of Refs.~\onlinecite{hermele04,moessner03}, were shown to possess a ${\rm U}(1)$ deconfined
phase, which supports gapped half-boson excitations, gapped ``magnetic
monopole'' topological defects, and a linearly dispersing photon mode with two
polarizations.  The low-energy theory is standard quantum electrodynamics with
massive electrically and magnetically charged scalar particles.  Models of
bosons on the square lattice with large ring exchange are in a different class
from those above, and also exhibit interesting physics.  Such models can support
an ``exciton Bose liquid'' phase, a two-dimensional analog of a Luttinger liquid
\cite{paramekanti02}, as well as nontrivial valence-bond solid (VBS) insulating
states \cite{sandvik02}.  Such states can undergo a direct quantum phase
transition to the superfluid \cite{senthil04}.

Despite much recent theoretical progress, clear experimental evidence for the
existence of exotic phases is still missing.  Furthermore,  relatively few
theoretical techniques exist to study such strongly correlated systems; perhaps
the most powerful to date is quantum Monte Carlo simulation, but the class of
models that can be productively studied is severely limited by the notorious
sign problem.  Atomic gases offer an alternative approach through the design of
\emph{quantum simulators}, where a microscopic Hamiltonian is implemented in a
quantum gas and its phase diagram is studied experimentally via controlling the
strength of the interaction terms.

\begin{center}
\begin{figure}[hbtp]
\begin{center}
\includegraphics[scale=0.3]{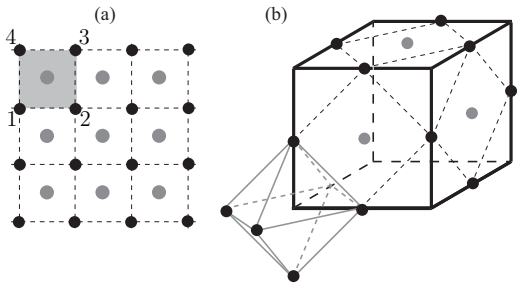}
\caption{ \label{2D-3Dsetup} (a) Two-dimensional setup: the bosons (black dots) 
are on the square lattice with the molecules (gray dots) in the center of each 
 plaquette. (b) Three-dimensional setup: the bosons (black dots) 
are on the links of the cubic lattice. Within each face there are four bosonic sites, 
which establish a plaquette (dashed square). The molecules (gray dots) 
are in the center of each plaquette.
   }
\end{center}
\end{figure}
\end{center}

In this Letter, we present the design of a ring exchange interaction for bosonic
gases subjected to an optical lattice. Such an atomic lattice gas is well
described by the Bose-Hubbard model \cite{jaksch98}
\begin{equation} 
  H_{\rs BH} = -J \sum_{\langle i, j\rangle} b^{\dag}_{i} b^{}_{j} +
  \frac{U}{2}\sum_{i} b^{\dag}_{i}b^{\dag}_{i} b^{}_{i}b^{}_{i} \label{BH},
\end{equation} 
where $U$ denotes the on-site repulsion and $J$ is the hopping energy with
$\langle i ,j \rangle$  denoting summation over nearest-neighbor sites.  The
additional ring exchange interaction involves four lattice sites forming a
square plaquette, and is driven by a resonant coupling of the bosons via a Raman
transition to a ``molecular'' two-particle state \cite{jaksch04}, see
Fig.~\ref{2D-3Dsetup}.  This state is subjected to an independent optical
lattice with its lattice sites at the plaquette centers. The symmetry of the
molecular state strongly influences the coupling; we are interested in a
$d$-wave symmetry of the molecule, which can be carried either by the relative
coordinate or the center of mass motion.  Then, the coupling to the molecular
state (created by $m_{{\rs \square}}^{\dag}$) takes the form
\begin{equation} 
  H_{\rs M} = \sum_{{\rs \square}} \nu m^{\dag}_{\rs \square} m_{{\rs
  \square}}^{ }  + g \sum_{{\rs \square}} \big[m^{\dag}_{\rs \square}
 \left(b_{1}b_{3}-b_{2}b_{4}\right)+ {\rm H.c.} \big]\text{.} \label{mHamiltonian}
\end{equation}
The summation runs over all plaquettes ${\rs \square}$.  The single-particle
states created by $b^{\dag}_i$ are called bosonic or atomic states to
distinguish them from the  ``molecular'' two-particle states.  Depending on the
setup, the atomic states reside either in the corners or on the edges of each
plaquette (see Fig.~\ref{2D-3Dsetup}), and are numbered counterclockwise. The
energy $\nu$ corresponds to the detuning from resonance, while $g$ is the
coupling strength determined by the Rabi frequency of the Raman transition.
While the Hamiltonian (\ref{mHamiltonian}) is interesting in its own right,
the connection to ring exchange is apparent upon integrating out the molecular
field perturbatively in $g/\nu$, which leads to the effective Hamiltonian
\begin{equation} 
  H_{\rs RE} = K  \sum_{{\rs \square}} \left(
  b^{\dag}_{1}b^{}_{2}b^{\dag}_{3} b^{}_{4} + b^{}_{1}b^{\dag}_{2} b^{}_{3}
  b^{\dag}_{4}\!- \!n_{1}n_{3}\! -\!n_{2}n_{4} \right), \label{ringexchange}
\end{equation}
with $K= g^{2}/\nu$. Note that the structure of the coupling in Eq.~(\ref{mHamiltonian}) 
also produces a next-nearest-neighbor interaction. 
The bosonic system turns metastable for large negative detuning.
However, the decay time easily exceeds typical experimental time scales of
atomic gases. Then, the perturbative expansion is again valid and allows for the
realization of a system with negative ring exchange interaction.


%

The design of the ring exchange Hamiltonian $H= H_{\rs BH}+ H_{\rs M}$ combines
standard tools for manipulating and controlling atomic gases:
First, we are interested in a cold atomic gas with two internal states, which are
subjected to independent optical lattices;  the internal states can be coupled
via a Raman transition.  Such a setup has been realized recently  using
spin-dependent optical lattices \cite{mandel03,duan03}. An alternative approach
is the trapping of alkaline earth metals, e.g., $^{88}$Sr,
where the first excited state $^{3}P_{1}$ exhibits a long life-time with a
different polarizability than the lowest energy state $^{1}S_{0}$
\cite{katori99}. Second, combining different laser configurations allows for the
design of complex lattice structures beyond the standard cubic lattice
\cite{santos04}, and third, the interaction between the atoms can be tuned by
magnetic or optical Feshbach resonances \cite{jaksch04}.

We introduce the notation  $\psi_{\rs a}({\bf x})$ and $\psi_{\rs b}({\bf x})$
for the field operators describing the two internal states. The microscopic
Hamiltonians then takes the form ($\alpha=a, b$)
\begin{displaymath} 
   H_{\alpha} \!=\! \int \! d{\bf x}\! \Big[ \psi_{\alpha}^{\dag}
   \Big(
   \frac{-\hbar^{2}\nabla^2}{2m}\!+\!V_{\alpha}\!+\!e_{\alpha}\Big)\psi_{\alpha}\!
   +\!\frac{g_{\alpha}}{2} \psi_{\alpha}^{\dag} \psi_{\alpha}^{\dag}  \psi_{\alpha} \psi_{\alpha}
   \Big] \label{GHamilton} 
\end{displaymath}
with $g_{\alpha}= 4 \pi \hbar^2 a_{\alpha}/m$ the interaction strength 
for scattering lengths $a_{\alpha}$. The
$e_{\alpha}$ are the homogeneous energy shifts between the internal states.  The
potential $V_{\alpha}({\bf x})$ accounts for an optical lattice driven by lasers
with wavevector $k=2\pi/\lambda$, 
with the strength $v_{\alpha}$ in
units of the recoil energy $E_{\rs r}= \hbar^2k^2/2m$.  The two internal states
are coupled via a Raman transition.  Transforming away the optical frequencies
within a rotating frame, the coupling takes the form
\begin{equation}
   H_{\rs R}= \hbar \Omega \int d{\bf x} \left[\psi^{\dag}_{\rs
   b}\psi_{\rs a}+ \psi^{\dag}_{\rs a} \psi_{\rs b}\right] 
\end{equation}
with $\Omega$ the Rabi frequency of the transition.

\begin{center}
\begin{figure}[hbtp]
\begin{center}
\includegraphics[scale=0.25]{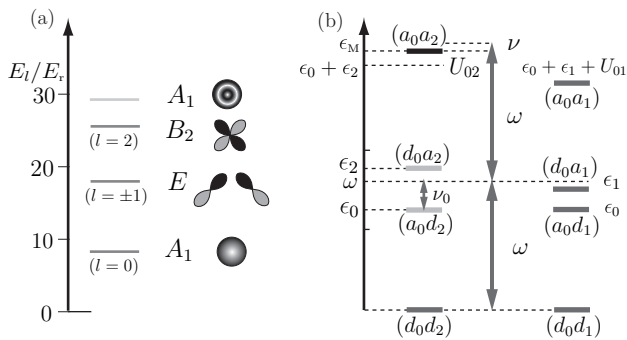}
\caption{ \label{energystructure} (a) Single-particle energies $\epsilon_{l}$ of the states
$a^{\dag}_{l,{\rs \square}}|0\rangle$ for $v_{\rs
a}=30$. (b) Energy levels of the two-particle states with symmetry $B_{2}$
and $E$. The frequency $\omega=
(\epsilon_{0}\!+\!\epsilon_{2}\!+\!U_{02}\!+\!\nu)/2 \hbar$ of
the Raman transition is chosen near resonance with the molecular state
$a^{\dag}_{0} a^{\dag}_{2}|0\rangle$, and far-detuned to the single-particle
excitations.  }
\end{center}
\end{figure}
\end{center}

We focus first on the two-dimensional setup shown in Fig.~\ref{2D-3Dsetup}a.
Confinement to two dimensions is achieved by a strong transverse optical
lattice, which quenches hopping between different planes. The remaining optical
lattice provides the square lattice structure for the atomic state $\psi_{b}$,
and takes the form $V_{\rs b}/E_{\rs r} = v_{\rs b} \left[\cos^2(k x/2)+
\cos^2(k y/2)\right]$. For  $v_{{\rs b}} \gtrsim 1$,  the mapping to the Bose
Hubbard model is well justified.  The optical lattice for the second internal
state $\psi_{\rs a}$, which is localized at the plaquette centers, takes the
form
\begin{displaymath} 
  \frac{V_{\rs a}}{E_{\rs r}} = v_{\rs a} \left\{ \left[\cos k
  x \! -\!\cos k y\right]^{2} 
  \!+\! \sin^{2}\left( k x/2\right)\!+\!  \sin^{2}\left( k y/2\right)\right\}.
\end{displaymath}
The first term is obtained by interference between standing laser waves along
the $x$- and $y$-directions, while the other terms represent a standard square
lattice.  The different lattice spacing of the two contributions is easily
achieved by a finite angle $2\pi/3$ between the interfering beams. We are
interested in a strong optical lattice $V_{\rs a}$, where tunneling between
different wells is quenched, and focus on the energy states within a single
well.  Then the structure of $V_{\rs a}$ produces strong shifts of the energy
states compared to those obtained  within a harmonic approximation. The states
are characterized by the irreducible representations of the symmetry group
$C_{4v}$ (\emph{i.e.} the point group of the square lattice); the low-energy
states and corresponding representations derived within a band structure
calculation  are shown in Fig.~\ref{energystructure} for $v_{\rs a}=30$.  The
state with energy $\epsilon_{l}$ in each plaquette ${\rs \square}$ is created by
the bosonic operator $a^{\dag}_{l, {\rs \square}}$ with $l=0,\pm1,2$. Of
particular interest is the state $|a_{2, {\rs \square}}\rangle$  corresponding
to the representation $B_2$, which transforms under $C_{4v}$ like the
polynomial~$x y$ (\emph{i.e.} $d_{x y}$ symmetry). In contrast to the harmonic
approximation, this state is non-degenerate.

For weak interactions, the Hamiltonian for the bosonic field $\psi_{\rs a}$ 
reduces to
\begin{equation}
  H_{\rs a} = \sum_{l,{\rs \square}} \nu_{l} a^{\dag}_{l, {\rs \square}} 
  a^{}_{l, {\rs \square}} + \sum_{l,l',{\rs \square}} \frac{U_{l,l'}}{2}  
  a^{\dag}_{l, {\rs \square}} a^{\dag}_{l', {\rs \square}} 
  a^{}_{l, {\rs \square}} a^{}_{l', {\rs \square}}
\end{equation}
with $U_{l,l'} \lesssim \epsilon_{l}$ the interaction energy, and
$\nu_{l}=\epsilon_{l}-\hbar \omega$ the energies of the excitations within the
rotating frame ($\omega$ is the frequency of the Raman transition).  The
coupling driven by the Raman transition takes the form $H_{\rs R} =\hbar \Omega
\sum_{l, {\rs \square}} \left[w_{l} a^{\dag}_{l,{\rs \square}} d_{l,{\rs
\square}} + {\rm H.c.}\right]$, with $\Omega$ the Rabi frequency.  Due to the
square symmetry, each operator $a_{l,{\rs \square}}$ couples to a special
structure of surrounding bosons.  The operators for which the coupling becomes
diagonal transform irreducibly under $C_{4 v}$ and are $d_{0,{\rs \square}}\sim
b_{1}+b_{2}+b_{3}+b_{4}$, $d_{\pm 1,{\rs \square}}\sim b_{1}\pm i b_{2}-b_{3}\mp
ib_{4}$, and $d_{2,{\rs \square}}\sim b_{1}-b_{2}+b_{3}-b_{4}$. The wave
function overlaps $w_{l}$ derive from the shape of the localized wave functions.
For typical parameters ($v_{\rs b} \approx 6$ and $v_{\rs a} \approx 30$) we
obtain $w_{l}\approx 0.1$.

We are interested in a setup with ``molecular'' two-particle states $|m_{\rs
\square}\rangle=m^{\dag}_{\rs \square}|0\rangle= a^{\dag}_{2,{\rs
\square}}a^{\dag}_{0, {\rs \square}}|0\rangle$ (with  $d_{x y}$ symmetry
$B_{2}$) resonantly coupled to the bosonic states
$b^{\dag}_{i}b^{\dag}_{j}|0\rangle$ with detuning $\nu$.  The energy of this
state is $\epsilon_{\rs M}= \epsilon_{0}+\epsilon_{2}+U_{02}$.  As the formation
of the molecule involves the virtual creation of a single particle excitation
$a^{\dag}_{l,{\rs \square}}|0\rangle$,  the frequency  of the Raman transition
is determined by $\omega=(\epsilon_{\rs M}+\nu)/2 \hbar$
(Fig.~\ref{energystructure}).  We require that the single-particle excitations
$a^{+}_{l, {\rs \square}}|0\rangle$ are far-detuned (\emph{i.e.}  $ w_{l}\hbar
\Omega < |\nu_{l}|$), and only the molecular states are resonantly coupled with
$ |\nu| \ll |\nu_{l}|$.  As the separation between the energy levels
$\epsilon_{l}$ is large ($\sim 2\sqrt{v_{\rs a}} E_{\rs r}$), this condition can
easily be satisfied for suitable interaction energy $U_{02}$.  Furthermore, it
may be of interest to tune the interaction energy via a Feshbach resonance to an
optimal value. For such strong interactions the molecular operator becomes
$m^{\dag}_{\rs \square}= \! c_1 a^{\dag}_{0,{\rs \square}} a^{\dag}_{2,{\rs
\square}} +  c_2 \big(a^{\dag}_{1, {\rs \square}}a^{\dag}_{1, {\rs \square}}+
a^{\dag}_{-1, {\rs \square}} a^{\dag}_{-1, {\rs \square}}\big)\! + \cdots$,
where the ellipsis denote admixture of higher energy states respecting the
$d_{x y}$ symmetry. The parameters $c_{1,2}$ and the energy $\epsilon_{\rs M}$
have to be determined from the solution of the two-particle problem within a
single well.

Integrating out the single-particle states $|l,{\rs \square}\rangle
=a^{\dag}_{l, {\rs \square}}|0\rangle$  perturbatively in $\hbar
\Omega/\nu_{l}$, we obtain the effective Hamiltonian $H_{\rs M}$, see
Eq.~(\ref{mHamiltonian}).  The last term in Eq.~(\ref{mHamiltonian}) accounts
for the coupling between molecules $|m_{\rs \square} \rangle$  and the atomic
states $| b_{i}\rangle$.  The operator $b_{1}b_{3}- b_{2} b_{4}$ is the only
second-order polynomial in $b_{i}$ transforming in the same representation
$B_{2}$ as the $d$-wave molecule $|m_{\rs \square}\rangle$. The coupling $g$ is
\begin{equation}
   g =-\hbar^{2} \Omega^{2} 
  \left[c_1 \: w_{0} w_{2}\left(\frac{1}{\nu_{0}}
  +\frac{1}{\nu_{2}}\right) + c_2 w_{1}^2\frac{1}{\nu_{1}} \right],
\end{equation}
which reduces to $g\sim 4 \hbar^{2} \Omega^{2} U_{0
2}/(\epsilon_{2}-\epsilon_{0})^{2}$ for weak interactions $U_{0 2} \ll
\epsilon_{0}$. We have dropped terms $\sim b_{i}^2$ as we assume that double
occupation of the bosonic sites is strongly suppressed by the on-site repulsion
$U$.  In addition to the Hamiltonian in Eq.~(\ref{mHamiltonian}), we obtain a
laser induced hopping term $\sum_{l, {\rs \square}} J_{l} d^{\dag}_{l,{\rs
\square}} d^{\vphantom\dagger}_{l,{\rs \square}}$ with $J_{l} = -  \hbar^{2}
\Omega^{2}|w_{l}|^{2}/\nu_{l}$.  The sign of this laser induced hopping depends
on the detuning. In principle it is possible to cancel these terms via
interference by an additional far-detuned Raman transition.  Furthermore, we
have  absorbed a  shift in the energy of the molecular energy into a
redefinition of $\omega$.  Tuning the Rabi frequency $\Omega$ sets the energy
scale of the coupling $g$, while the detuning $\nu$ is controlled by the
frequency $\omega$ of the Raman transition; this allows the system to be tuned
through a resonance.

The zero temperature phase diagram of the two-dimensional setup shown in
Fig.~\ref{2D-3Dsetup}a  with the Hamiltonian $H= H_{\rs BH}+ H_{\rs M}$ has not
yet been studied.  However, by considering appropriate limits we suggest there
is potential for interesting physics in the intervening regime.  We let $q$ be
the average number of atoms per unit cell (\emph{i.e.} molecular states counted
twice),  and consider the filling $q = 1/2$.  For large  positive detuning
$\nu\gg g$ with $J \gg g^2 / \nu$, the system reduces to the conventional
Bose-Hubbard model, see Eq.~(\ref{BH}), and the bosons establish a superfluid
phase due to the incommensurate filling.  In the opposite limit of large
negative detuning ($|\nu| \gg J, g$ and $\nu < 0$), all bosons are paired into
molecules, and we can think in terms of an effective molecular Bose-Hubbard
model at 1/4 filling.  For $J \ll g$, perturbation theory in $g/|\nu|$ generates
a nearest- and next-nearest-neighbor repulsion between molecules $U_{\rs M} \sim
g^4 / |\nu^3|$.  Furthermore, a molecular ring exchange term is generated at the
same order.  The ground state of the resulting model is not known, but is likely
to be a molecular charge density wave with $\langle m^{\dag}_{\rs \square}
m^{\vphantom\dagger}_{\rs \square} \rangle$ larger on every other row and
column.  There is no difference at the level of symmetry between this state and
a ``plaquette'' valence bond solid (VBS) of the atomic bosons.  We therefore
suggest that this system is a candidate for a deconfined quantum critical point
\cite{senthil04} as the system is tuned between the VBS and the bosonic
superfluid.

The above design of a ring exchange interaction for bosonic systems is a
building block that can be applied to many different setups and lattice
structures \cite{futurepaper}.  Of particular interest are models exhibiting a
local gauge invariance that may allow for the realization of deconfined
insulators. 

In the following, we focus on the three-dimensional setup shown in
Fig.~\ref{2D-3Dsetup}b, that can likely access such a ${\rm U}(1)$ deconfined
state. The lattice structure is described by a cubic lattice with the bosonic
states on the links.  Each cubic face forms a plaquette involving four link
bosons created by the operator $b^{\dag}_{ij}$, and the molecules are again 
placed at the plaquette centers, \emph{i.e.}
in the center of each cubic face.  The lattice of bosonic states can also be
viewed as the lattice of corner-sharing octahedra with their centers at the
cubic sites.  With this definition of the bosonic and molecular sites, we can
again derive the Hamiltonian $H=H_{\rs BH}+ H_{\rs M}$ following the procedure
discussed above for the two-dimensional setup \cite{futurepaper}. We expect that
the phase diagram of this system is dominated by a superfluid phase for large
hopping, while in the limit of small hopping more exotic states may result. To
better understand the possibilities, we consider the limit of vanishing hopping
$J=0$.  For large detuning $|\nu| \gg g$, the molecules can be integrated out to
obtain the cubic ring exchange model of Ref.~\cite{hermele04}, with an
additional interaction term, see Eq.~(\ref{ringexchange}).  (Note, that a
unitary transformation allows one to change the sign of the ring exchange term.)
This model is a ${\rm U}(1)$ lattice gauge theory and, at least over some region
of the parameter space, it can enter its Coulomb phase, {\em i.e.}  a ${\rm
U}(1)$ deconfined insulator \cite{hermele04}.  It has been shown that this
state is stable to \emph{all} perturbations, including those that break the
gauge invariance such as a boson hopping $J$. Then, in the
presence of a small but finite hoppping $J$, the gauge
structure goes from an explicit, microscopic property to an emergent one present
only in the low-energy theory.

Remarkably, the three-dimensional setup shown in Fig.~\ref{2D-3Dsetup} with 
the Hamiltonian $H=H_{\rs BH}+ H_{\rs M}$ and quenched hopping $J=0$ 
is even  a ${\rm U}(1)$ lattice
gauge theory for arbitrary $g/\nu$.  This can be seen by considering a cubic
site $i$, letting $L(i)$ be the 6 cubic links  and $P(i)$ the 12
plaquettes containing $i$.  Then the local object $G_i = \sum_{ij \in L(i)}
b^{\dag}_{ij} b^{\vphantom\dagger}_{ij} + \sum_{{\rs \square} \in P(i)}
m^{\dag}_{{\rs \square}} m^{\vphantom\dagger}_{{\rs \square}}$ is a conserved
${\rm U}(1)$ ``gauge charge.''  A straightforward ``spin wave'' treatment allows
one to write down a low-energy theory of liquid phases in ring exchange models
\cite{paramekanti02, hermele04}, and we obtain here an artificial photon mode
with two polarizations. This analysis demonstrates that this model is likely to
support a ${\rm U}(1)$ deconfined insulator, which is likely to be continuously
connected to the ${\rm U}(1)$ deconfined phase discussed above and in
Ref.~\onlinecite{hermele04} for the large-$\nu$ limit.  We also note that $H =
H_{\rs BH} + H_{\rs M}$ should be amenable to quantum Monte Carlo simulation
when $J=0$, as a simple unitary transformation renders all matrix elements of
$e^{-\tau H}$ nonnegative in the number basis.

Finally, we remark that the $d$-wave molecular state couples to a special
structure of the surrounding bosons.  In addition to the applications discussed
above for the \emph{design} of strongly correlated models, it should be possible to use
the coupling of this state as a \emph{probe} of this structure within phases of
the single particle system.  This may provide a powerful tool for the detection
of unconventional order in strongly correlated systems.

\acknowledgments{We are grateful to A.\ Muramatsu for helpful discussions.  This
research is supported by the Austrian Science Foundation and the Institute of
Quantum Information (H.P.B. and P.Z.\ ), and NSF Grant Nos. DMR-0210790 and
PHY-9907949 (M.P.A.F.\ and M.H.\ ).}

\bibliographystyle{../../../Refbib/apsrev}
\bibliography{../../../Refbib/journals,../../../Refbib/ref}
 
\end{document}